\documentclass[aps,llncs, twocolumn]{revtex4-1} 
\usepackage{graphicx}
\usepackage{amsmath}
\usepackage{float}
\begin{document}
\title{Enhanced, homogeneous type II superconductivity in Cu-intercalated PdTe$_2$} 

\author{Aastha Vasdev$^1$, Anshu Sirohi$^1$}

\author{M. K. Hooda$^2$, C. S. Yadav$^2$}

\author{Goutam Sheet$^1$}
\email{goutam@iisermohali.ac.in}
\affiliation{$^1$Department of Physical Sciences, 
Indian Institute of Science Education and Research Mohali, 
Mohali, Punjab, India}

\affiliation{$^2$School of Basic Sciences, Indian Institute of Technology Mandi - Mandi-175005 (H.P.) India}
 
\begin{abstract}

Though the superconducting phase of the type-II Dirac semimetal PdTe$_2$ was shown to be conventional in nature, the phase continued to be interesting in terms of its magnetic properties. While certain experiments indicated an unexpected type-I superconducting phase, other experiments revealed formation of vortices under the application of magnetic fields. Recently, scanning tunneling spectroscopy (STS) experiments revealed the existence of a mixed phase where type-I and type-II behaviours coexist. Here, based on our temperature and magnetic field dependent STS experiments on Cu-intercalated PdTe$_2$ we show that as the critical temperature of the superconducting phase goes up from 1.7 K to 2.4 K on Cu-intercalation, the mixed phase disappears and the system becomes homogeneously type-II. This may be attributed to an averaging effect caused by quasiparticle exchange between type-I and type-II domains mediated by the Cu atoms and to decreased coherence length due to increased disorder.

\end{abstract}
\maketitle


The candidate type II Dirac semimetal PdTe$_2$ \cite {Noh, Fei,Amit2} superconducts below 1.7 K \cite {leng,Guggenheim,Kjekshus,
Raub,Alidoust}. Though initial experimental papers suggested the possibility of an unconventional nature of the superconducting phase of PdTe$_2$, a number of experiments confirmed that the superconductivity in PdTe$_2$ is conventional in nature \cite{shekhar,Amit}. However, some of these experiments found evidence of type I behaviour as far as the magnetic properties of the superconducting phase of PdTe$_2$ are concerned \cite{leng,Shapiro,Salis}. These contradicted with certain other experiments where the magnetic properties were found to be consistent with type II superconductivity \cite{Clark}. More recently, magnetic field dependent scanning tunneling spectroscopy experiments revealed coexistence of type I and type II domains in PdTe$_2$ where such an  inhomogeneity of the superconducting phase was attributed to electronic inhomogeneities observed in the normal state of PdTe$_2$ \cite{anshu}. It turned out that the electronic inhomogeneity could alter the local electron density and the local coherence length thereby leading to a distribution of the Ginzburg-Landau-Abrikosov-Gorkov (GLAG) parameter ($\kappa$) over the surface of PdTe$_2$ \cite{Abrikosov}. Within this picture, it should be possible to disorder the system so that inter-domain scattering of the quasiparticles can average out the $\kappa$ distribution and make the system homogeneous. In this paper, we report the observation of such averaging effect in Cu-PdTe$_2$ by magnetic field dependent scanning tunneling spectroscopy experiments. The superconducting phase becomes homogeneously type II while the critical temperature gets enhanced at randomly distributed domains on the surface due to random distribution of the intercalates .   

The effect of Cu-intercalation on the electrical and thermal transport properties in PdTe$_2$ was investigated in detail in the past \cite{Ryu,Yan}.
 Such measurements on polycrystalline samples indicated that the higher $T_c$ in Cu-PdTe$_2$ is due to enhanced electron-phonon coupling parameter ($\lambda_{ph}$) and increased density of states upon Cu-intercalation\cite{Hooda}. For the spectroscopic measurements presented in this paper we used high quality single crystals of Cu-intercalated PdTe$_2$ with 4\% Cu in the crystals. The shiny single crystals were first mounted in a ultra-high vacuum (UHV) cleaving stage integrated with the STM. The crystals were then cleaved at 80 K using a cleaver attached to an UHV manipulator. After cleaving, the crystals were immediately transferred to the STM head which is cooled down to 310 mK using a H$^3$-based cryostat. The sample is placed at the center of a solenoidal superconducting magnet with a maximum field strength of 11 Tesla. All the STM/S experiments reported in this paper were performed with sharp metallic tips of tungsten (W). The tips were fabricated by electro-chemical etching and were cleaned by electron beam bombardment under UHV in a preparation chamber attached to the STM prior to the low temperature experiments. The formation of the superconducting phase with $T_c$ = 2.6 K is seen in the susceptibility versus temperature data presented in Figure 1(b).    

In Figure 1(a), we show an atomic resolution image of the surface of Cu-PdTe$_2$. A careful inspection of this image reveals two types of defects -- (i) triangular defects as in PdTe$_2$ and typical for chalcogenides and (ii) dark spots surrounded by well defined patterns. While (i) were seen in non-intercalated PdTe$_2$, (ii) are new and therefore originate from the intercalated Cu atoms. Clear image of the lattice captured from a defect-free region is shown in the $inset$ of Figure 1(a). The patterns formed around the Cu atoms could be an
 interference pattern caused by incoming quasiparticles and their counterparts scattered from the Cu atoms. We have investigated the energy dependence of such patterns in detail and found that the pattern does not disperse with energy. Hence, the states participating in formation of the patterns are most likely topologically trivial surface states.

The scanning tunnelling spectroscopy revealed two types of spectra. One type captured at 310 mK is shown as a blue line in Figure 1(c). The coherence peaks are clearly seen below which the differential conductance becomes zero indicating formation of a full gap \cite{Teknowijoyo}, as expected for  BCS superconductor. The spectrum is fitted using Dyne's equation given by $N_s(E) = Re\left(\frac{(E-i\Gamma)}{\sqrt{(E-i\Gamma)^2-\Delta^2}}\right)$, where $N_s(E)$ is the density of states at energy $E$, $\Delta$ is the superconducting energy gap and, $\Gamma$ is an effective broadening parameter incorporated to take care of slight broadening of the BCS density of states possibly due to finite quasi-particle life time \cite{Dynes}. For the type of spectra shown in Figure 1(c), the extracted value of $\Delta$ is 0.53 meV. To note, the experimental spectra slightly deviate from the theoretical fits just above the coherence peaks -- this deviation was not seen in case of non intercalated PdTe$_2$. As we will discuss later, this type of spectra correspond to a higher $T_c \sim$ 2.4 K. At certain other points, the spectra are seen to be different and their analysis reveal a smaller $\Delta \sim$ 0.44 meV. These spectra also correspond to a lower $T_c \sim$ 1.7 K. The deviation of the experimental data from the theoretical fits are seen for these spectra as well. For the spectra with smaller $\Delta$, though the $T_c$ remains to be same as that of pure PdTe$_2$, the gap is measured to be higher than in pure PdTe$_2$ ($\sim$ 0.32 meV)\cite{shekhar}. Hence, the the effect of intercalation is present everywhere including the points where a lower $T_c$ (and lower $\Delta$) is measured.

\begin{figure}
		\includegraphics[width=.5\textwidth]{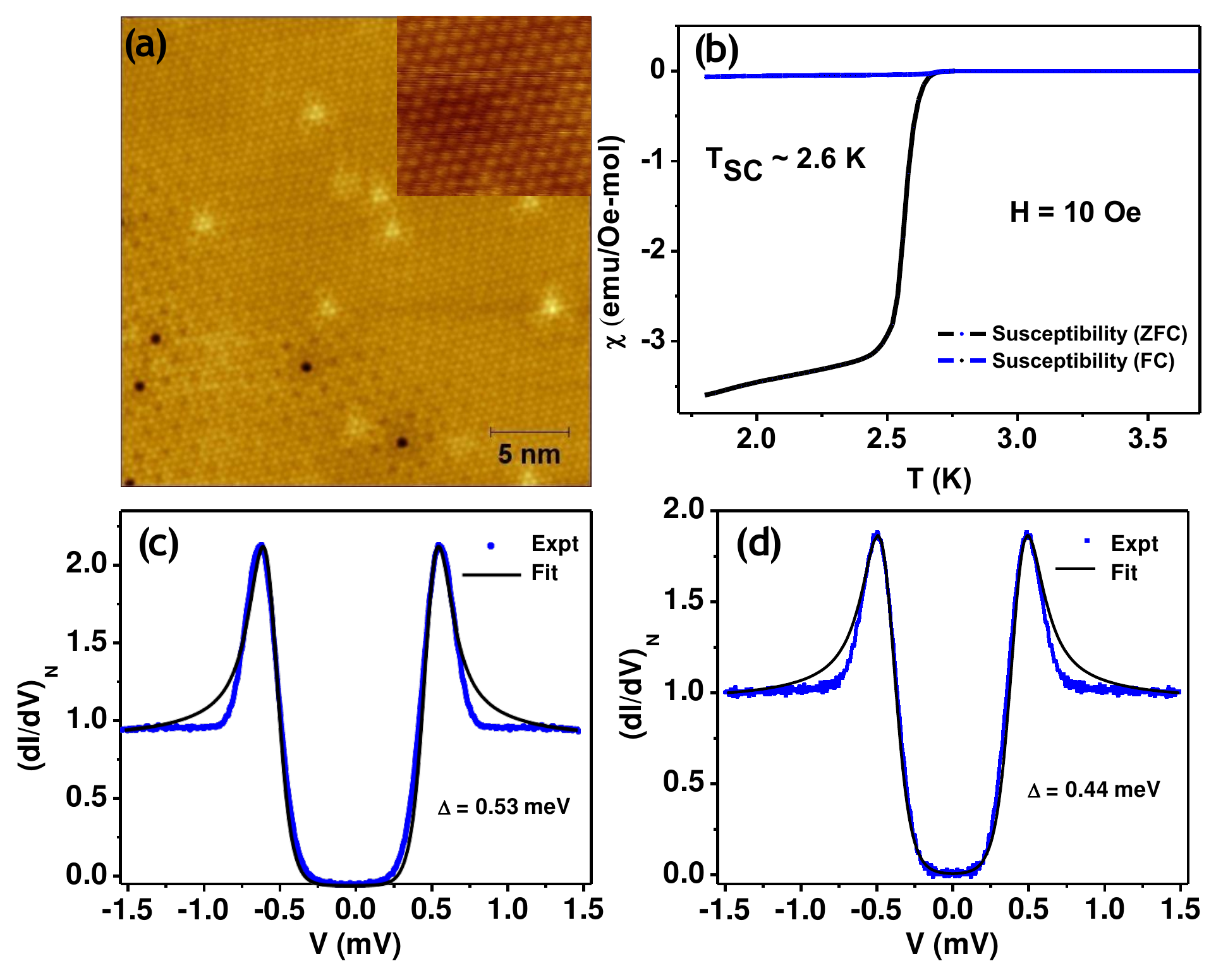}
		\caption{(a) Atomic resolution image (30 nm x 30 nm) of the Cu-PdTe$_2$ surface. (b) Susceptibility as a function of temperature (both ZFC and FC). (c,d) two representative STS spectra with best theoretical fits}

\end{figure}

From a comparative consideration of the results discussed above, it appears that the superconducting phase with higher $T_c$ that is expected to emerge upon Cu-intercalation (as per bulk measurements) appears at several places (as in figure 1(c)) while, at several other places, the enhancement of superconducting properties is less (as in Figure 1(d)). This suggests that the phase with higher $T_c$, though does not arise uniformly throughout the crystals, exceeds the percolation threshold thereby contributing predominantly in bulk transport measurements. In order to search for a correlation between the domains with higher $T_c$ phase and the defect states due to Cu, we performed local spectroscopy near and away from such defects. It was observed that the higher $T_c$ phase appears more frequently in the Cu rich regions.

\begin{figure} [h!]
		\includegraphics[width=.5\textwidth]{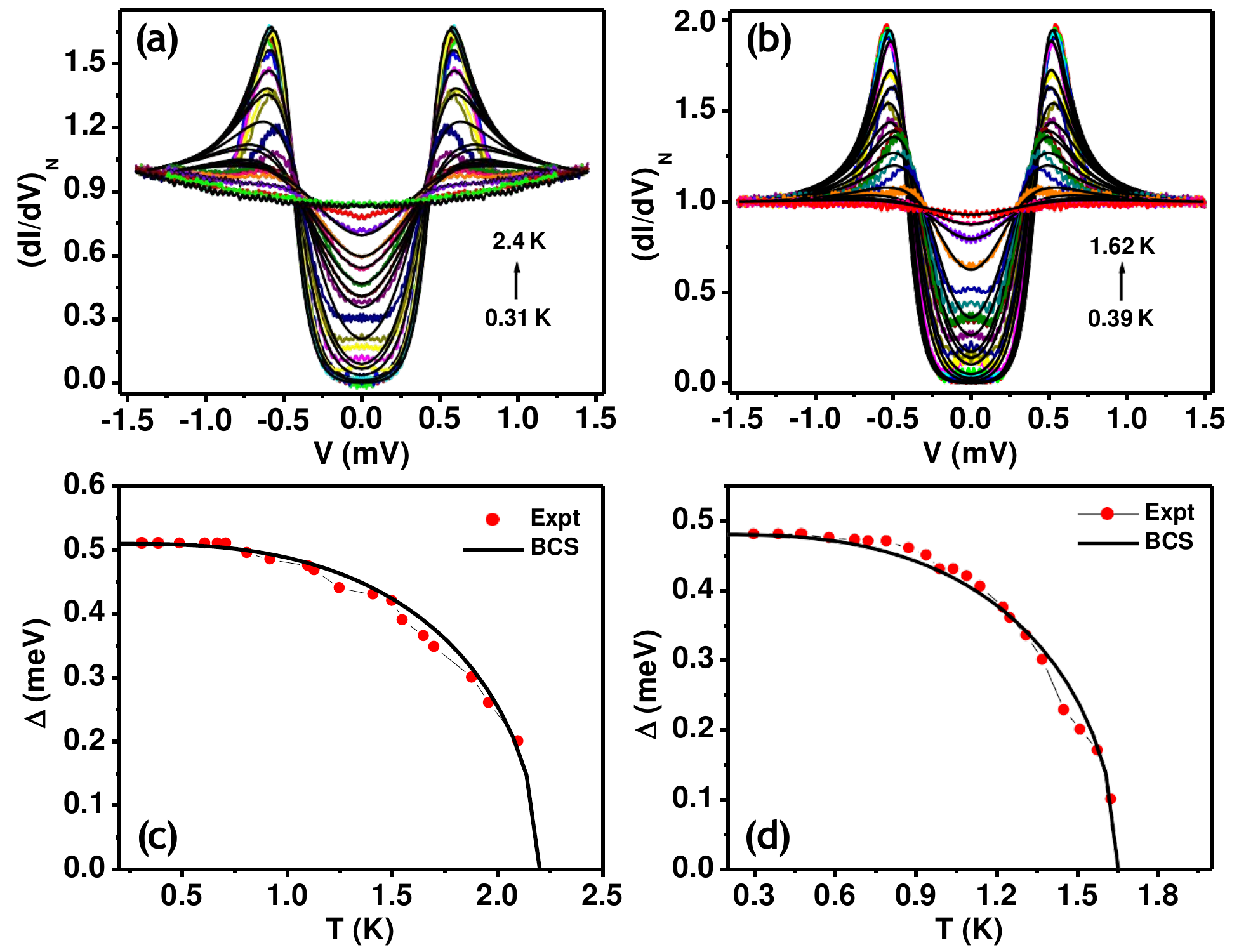}
		\caption{ Temperature dependence of conductance spectra  (a) with higher T$_c$ (b) with lower T$_c$   (c, d) $\Delta$ vs $\Gamma$ plot
extracted from (a, b) respectively.}	

\end{figure}

One of the most intriguing features of the superconducting phase of PdTe$_2$ is that despite hosting topologically non-trivial bands crossing the Fermi surface\cite{Zheng,Huang}, the superconducting order parameter is found to be conventional and is seen to follow BCS-like temperature dependence\cite{Clark,Kyoo}. The conventional nature of the temperature dependence is also found in case of the two energy gaps that are measured in Cu-intercalated PdTe$_2$. The temperature dependent spectra along with the theoretical fits are shown in Figure 2(a) for the phase with higher $\Delta$ and in Figure 2(b) for the phase with lower $\Delta$. Surprisingly, in both the cases, while the lowest temperature spectra showed no additional features other than the sharp coherence peaks, at slightly elevated temperatures a small zero-bias conductance peak (ZBCP) is also seen to appear. When the emergence of the ZBCP is considered along with the slight deviation of the experimental spectra from the theoretical fit, it is important to first investigate whether a tiny unconventional component in the order parameter can be present here. Such a ZBCP may appear due to a small unconventional component in the gap function that has a different temperature dependence than the dominating conventional component. In order to understand this point clearly, let us consider a temperature dependent gap amplitude given by $\Delta(T) = \alpha \Delta_{s} (T) + \beta \Delta_{un} (T)$, where $\Delta_s$ is the conventional component and $\Delta_{un}$ is the unconventional component of the gap, and $\alpha$ and $\beta$ are the respective coefficients. Now, since the ZBCP does not appear at low temperatures, it is clear that at those temperatures, $\alpha \Delta_s \gg \beta \Delta_{un}$. Now, with increasing temperature, due to the possible difference in temperature dependence of $\Delta_s$ and $\Delta_{un}$, at some point, the contribution of $\Delta_{un}$ may become comparable to $\Delta_s$ thereby making the unconventional contribution detectable in the form of the ZBCP. Beyond this temperature, as shown in the data, all the spectroscopic features related to superconductivity get systematically reduced with a complete disappearance of the spectral features at $T_c$. It should be noted that while the possibility of an unconventional component cannot be ruled out, the ZBCP may also originate from other sources, like inelastic impurity scattering with non-trivial temperature dependence. Since no ZBCP is seen in pure PdTe$_2$, here, the enhanced scattering may be due to the impurity states of the Cu intercalates. Since the strength of the ZBCP is significantly smaller than the overall spectra, we have fitted the temperature dependent spectra ignoring the ZBCP. The temperature dependent $\Delta$ thus obtained, are plotted in Figure 2(c) and Figure 2(d) respectively for the larger $\Delta$ and the smaller $\Delta$. The red dots represent the measured values of $\Delta$ and the solid lines show the expected temperature dependence within BCS theory, barring a slight deviation at higher temperatures. In both the cases, the temperature evolution of the spectra is consistent with BCS theory\cite{Bardeen}. From the measured valued of $\Delta$ and $T_c$, it is seen that $\Delta/k_BT_c$ are $\sim$ 2.56 and 3.15 corresponding to the larger $\Delta$ and the smaller $\Delta$ respectively. This means, the regions where the effect of intercalation is smaller, $\Delta/k_BT_c$ is higher. This shows that the elctron-phonon coupling strength decreases upon Cu-intercalation. This result apparently contradicts the theoretical finding of an enhanced electron-phonon coupling constant ($\lambda_{ph}$) in Cu-intercalated PdTe$_2 $\cite{Hooda}. Therefore, our results suggest that the enhancement in $T_c$ in Cu-PdTe$_2$ is primarily due to the enhanced density of states at the Fermi energy and not due to altered electron-phonon coupling. However, these two processes may compete with each other giving rise to an effective enhancement of superconducting properties in Cu-PdTe$_2$. 

\begin{figure} [h!]
		\includegraphics[width=.5\textwidth]{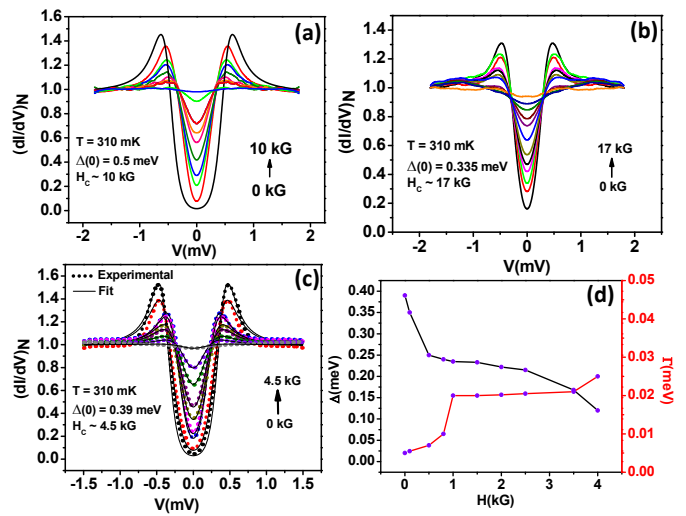}
		\caption{ Normalized STS data at three different points with varying magnetic fields  upto (a)10 kG (b) 17 kG (c) 4.5 kG. Colored dots show experimental data points and the black lines show theoritical fits using Dyne's equation   (d) $\Delta$ and $\Gamma$ vs H plot
extracted from (c)}	

\end{figure}

Now we focus on the magnetic field dependent spectroscopic measurements on Cu-intercalated PdTe$_2$. As it was discovered before, the surface of pure PdTe$_2$ hosts co-existence of domains with type I and type II superconductivity. In the type I domains, the superconducting spectra suddenly disappeared under the application of a low magnetic field $\sim$ 250 Oe. In the type II regions, the spectra survived up to much higher magnetic fields and the spectral features were seen to disappear continuously. Such variation of the type of superconductivity based on magnetic properties could be attributed to inhomogeneity of local coherence length ($\xi$). An inhomogeneous background possibly responsible for such inhomogeneity in $\xi$ was found in the normal state conductance maps. Within this understanding, introduction of disorder into the crystal is expected to reduce $\xi$ everywhere. 

In Figure 3(a,b,c) we show magnetic field dependent tunnelling conductance spectra captured at 310 mK at three different points on the surface of Cu-PdTe$_2$. In all cases, the conductance spectra, regardless of the corresponding superconducting energy gaps and $T_c$, evolved systematically and smoothly with magnetic field. The spectral features continuously became less prominent eventually disappearing at a critical field characteristic for the point where the spectroscopic measurement was performed. The critical fields measured in these cases are 10 kG, 17 kG and 4.5 kG respectively. The ubiquitously observed very low critical field ($\sim$ 220 Oe) in PdTe$_2$ was not detected on any point in Cu-PdTe$_2$. It should be noted that the measured critical fields at different points is much larger than the expected "surface critical field" ($H_s\sim$ 1.7$H_c$) in BCS superconductors\cite{Saint,Saint2,Tinkham}. Hence the enhancement is not due to a naturally enhanced surface critical field. In Figure 3(c), we also show Dyne's fit of one set of the magnetic field dependent spectra and the variation of the superconducting energy gap with increasing magnetic field is shown in Figure 3(d), where the smooth disappearance of the gap is clearly visible. This observation indicates that the superconductivity of Cu-PdTe$_2$ has become uniformly of type II throughout the crystal upon intercalation of Cu. This is different from the mixed type I and type II superconducting phase that was observed in pure PdTe$_2$ \cite{anshu}.

The observation of the uniformly type II behaviour of the superconducting phase upon Cu-intercalation supports the idea that the mixed type I and type II phase in pure PdTe$_2$ originated from a distribution of the local coherence length ($\xi$) and consequent distribution of $\kappa$. Introduction of disorder in the lattice causes increased scattering and a consequent decrease in the electronic mean free path ($l$) leading to a decreased $\xi$ following the relation $1/\xi = 1/\xi_0 + 1/l$, where $\xi_0$ is the BCS coherence length in the absolute clean limit. When the local $\xi$ in the type I regions decrease, the GLAG parameter ($\kappa$) also decreases and consequently these regions should show magnetic properties of that of type II superconductors. The criteria for the transition from a type I to a type II superconducting state in terms of the mean free path ($l$) has been discussed elsewhere where it was also shown that beyond a critical value of $l$, the superconducting phase is expected to become homogeneously type II\cite{anshu}.

Another aspect of the magnetic field dependence of $\Delta$ that needs attention is a sudden change in $\Delta$ at $H\sim$ 100 Oe, where $\Delta$ decreases to almost 50 \% of the zero field value. In a type II superconductor, such a sudden change may be associated with the emergence of a superconducting vortex in an area very close to the STM tip thereby causing the suppression of local $\Delta$. As shown in the $\Gamma$ vs. $H$ plot in Figure 3(d), as $\Delta$ shows a sudden drop, $\Gamma$ shows a simultaneous upward jump (by a factor of 6) at approximately the same magnetic field. $\Gamma$ represents pair breaking mechanisms leading to finite quasiparticle lifetime. Hence, this observation is consistent with the field-induced pair breaking under the STM tip, possibly due to the emergence of a nearby vortex core. However, despite our repeated attempts, we did not find spectra with clear zero-bias conductance peak in presence of magnetic field. To note, such zero-bias peaks in presence of sufficiently large magnetic field were frequently observed in pure PdTe$_2$. It is possible that in case of Cu-PdTe$_2$, the nature of the vortex core spectrum changes and becomes featureless and hence remains undetectable in our experiments. Beyond the field at which the sudden change happens, both $\Delta$ and $\Gamma$ change gradually and beyond a certain field $\Gamma$ becomes larger than $\Delta$ thereby suppressing the  key spectroscopic features. 

In conclusion, we have performed detailed scanning tunnelling microscopy and spectroscopy on Cu-intercalated crystals of PdTe$_2$ at different temperatures and magnetic fields. The motivation was to address the fate of the intriguing magnetic properties of PdTe$_2$ upon Cu-intercalation. We found that Cu-intercalation enhances the critical temperature throughout the crystal surface. In addition to $T_c$ enhancement, due to increased disorder, the surface of Cu-intercalated PdTe$_2$ becomes uniformly type II with no signature of co-existing type I phase anywhere. The effective critical field at all points becomes higher than that in the type I phase of pure PdTe$_2$. The results support the idea that the magnetic properties and electronic structure in PdTe$_2$ surface are highly correlated and the peculiar magnetic properties reported before are a consequence of such correlation.

AV thanks UGC for senior research fellowship (SRF). GS thanks financial support from a research grant (grant number: \textbf{DST/SJF/PSA-01/2015-16}) under Swarnajayanti Fellowship awarded by the Department of Sceience and Technology, Govt of India.


\end{document}